\documentclass[manuscript]{copernicus}  

\nolinenumbers
\usepackage{color}
\usepackage[T1]{fontenc}
\usepackage{amsmath}

\begin{document}

\title{Diffuse Josephson Radiation in Turbulence
}

{\author[1,3]{R. A. Treumann}
\author[2]{Wolfgang Baumjohann$^*$}
\affil[1]{International Space Science Institute, Bern, Switzerland}
\affil[2]{Space Research Institute, Austrian Academy of Sciences, Graz, Austria}
\affil[3]{Geophysics Department, Ludwig-Maximilians-University Munich, Germany\protect\\
$^*${(Correspondence to: Wolfgang.Baumjohann@oeaw.ac.at)}
}

}

\runningtitle{Josephson radiation}

\runningauthor{R. A. Treumann \& Wolfgang Baumjohann}

\received{ }
\pubdiscuss{ } 
\revised{ }
\accepted{ }
\published{ }


\firstpage{1}

\maketitle

  

\noindent\textbf{Abstract}.-- 
{The possibility of generating diffuse radiation in extended astronomical media by plasma turbulence is investigated under the assumption that the turbulence can be understood as an ensemble of small-scale magnetic filaments (narrow current sheets) forming a texture around a large number of magnetic depletions (voids). On astronomically microscopic scales the dilute high temperature medium (plasma) is to be considered ideally conducting forming a collection of Josephson junctions between two such adjacent quasi-superconductors. The oscillation frequency of those junctions depends on the part of the spectrum that contributes to the oscillation causing weak {radio backgrounds}. Lowest Josephson frequencies/energies near zero may become sources of quasi-stationary magnetic fields.  } 

\section{Introduction}
Diffuse radiation from extended astrophysical objects like clusters of galaxies is conventionally attributed to synchrotron radiation \citep{jackson1975,rybicki1979} from a distribution of relativistic particles which have been accelerated by some diffusive Fermi-like acceleration mechanism \citep[cf., e.g.,][]{schlickeiser2002} in the assumed always present magnetohydrodynamic plasma turbulence, both well-established, commonly accepted and successfully applied processes which provide valuable information about the physical state of the radiation sources, in particular the energy of the radiating particles and the strength of the scattering magnetic fields. 

Here we propose a different mechanism which in some cases may add to provide additional diagnostic information about turbulent emission sources. It is not based on the assumption of an energetic particle distribution but restricts to the presence of turbulence in extended objects like, for instance, turbulent supernova remnants, galaxies, clusters of galaxies, and possibly even the cosmological large-scale structure of the universe which exhibits a particular texture consisting of filamented matter and voids. The weak emission generated may in such large extended turbulent media sometimes add up to observable intensities. 

In configuration space  turbulent media are not smooth on the mesoscopic  and microscopic scales but consist of a very large number of vortices which on the large scale appear about homogeneous. Interaction on the small scales provides effects which could map into observations. A well known example are magnetohydrodynamic instabilities (mirror modes, alfv\'enic structures, discontinuities, shock waves, etc.) which structure any extended turbulent plasmas. The medium subject to this kind of turbulence consists of magnetic vortices, small-scale current sheets, and magnetic depletions separated by on the larger scale very narrow magnetic walls and filaments which, in well developed turbulence, may form more or less irregular chains of magnetic voids. 

Chains of this kind and irregular distributions of partially depleted voids in the turbulent medium, which on those scales is almost perfectly conducting, can if forming magnetic voids be interpreted as a network of Josephson junctions. These are not connected to form a large multi-junction array, rather they can be considered as a multitude of single junctions distributed over the entire turbulent volume. Any interaction occurs only between nearest neighbours while rapidly decreasing with distance. Each individual junction then consist of just two neighbouring magnetic depletions which are connected via the narrow separating magnetic wall or filament, structures which belong to the turbulent texture that is generated in the volume by the free energy source of the turbulence: active galactic nuclei (AGNs), supernova remnants (SNRs), or any other object/process responsible for feeding turbulence like for instance collisionless shocks \citep{balogh2013} and their environments \citep{eastwood2005,lucek2005}, the interstellar medium \citep{haverkorn2013} or stellar winds like the example of the solar wind  \citep{goldstein2005,goldstein1995,khabarova2013} shows.

\section{Josephson junctions}
The physics of such junctions had been discovered and formulated sixty years ago \citep{josephson1962,josephson1964} and in quantum devices has become an extraordinarily important diagnostic tool for measuring tiny electric potential differences. For natural systems it has been reviewed in recent work focussing on meso-scale mirror mode turbulence \citep{treumann2021} in near-Earth space. The scales will be vastly different, but it is reasonable to assume that turbulence basically forms structures occupying the range from the largest mechanically driven alfv\'enic scales down into the ion inertial scale range in interaction with their nearest neighbours, preferentially if of similar scale, by exchanging tunnelling currents \citep{josephson1965,bogoliubov1958,valatin1958} across the separating magnetic filamentary walls. {Their magnetic effect is the classical skin depth which allows the magnetic field to penetrate a short distance into the matter.} 

In the semi-classical approximation these currents are real electric currents indeed, flowing perpendicular to the separating magnetic fields. These currents are carried solely by electrons. They temporarily bridge the wall between the voids but are not allowed to penetrate the void over more than a microscopic skin depth $\lambda_e=c/\omega_e$ (with $\omega_e^2=e^2N_0/\epsilon_0 m_e$ the squared plasma frequency, $m_e$ electron mass, and $N_0$ the ambient density) which implies that they become reflected and oscillate back and forth while being locally confined to the filaments. The oscillation frequency of the fluctuating currents is high, the order of the Josephson frequency 
\begin{equation}
\nu_J\equiv\frac{\omega_J}{2\pi}= \frac{|q|}{2\pi\hbar}\langle\Delta V\rangle \approx 2.5\times10^{5}\langle\Delta V\rangle \quad\mathrm{GHz}
\end{equation}
{(with $q=-e$ electron elementary charge)} if only a weak electric potential difference $\langle\Delta V\rangle$ (in Volts) is applied to the junction \citep[cf., e.g.][or the above cited original publications]{fetter1971,ketterson1999}.  Clearly, even for small electric potentials this frequency is high, {actually far above any cyclotron frequency $\omega_J\gg\omega_{ce}=eB/m_e$ in the ambient magnetic field $\mathbf{B}$ such that the electron magnetic moment $\mu_e=T_e/B$ is not conserved. This violates adiabaticity and permits tunnelling.} The source of the potential can either be found in the always present thermal fluctuation level of plasma turbulence \citep{krall1973,baumjohann1996} or in the turbulent streaming itself {as we will demonstrate below}. 

Plasmas are quasi-neutral, and any potential difference across the void-separating walls imposed by the turbulent flow will necessarily be rather small locally, because the walls are narrow and the cross-potential drop is small. This holds in particular in weak magnetic fields where {in the near Earth space and solar wind, for instance} the average flow-electric fields typically are of the order of $\langle \delta E\rangle\sim$ few $\mu$V\,/\,m, becoming at most mV\,/\,m which, for narrow junction boundaries of width, say, $L_n\lesssim 10\,\lambda_e$, may shift the Josephson frequency down into the (astronomically interesting) radio frequency range $\sim 1$ GHz.\footnote{{When using numbers and dimensions from the magnetosheath or solar wind we refer to \citep[][]{lucek2005,goldstein2005}, otherwise, when having in mind astrophysical applications in clusters of galaxies the appropriate references are \citep[][]{walker2019,simionescu2019}.}} Larger potentials, which are not necessarily expected to occur, as also broader and therefore probably less effective walls would shift the frequency up. Such strong electric fields are barely expected {except locally in collisionless shock transitions or in relativistic streams} and, presumably, become depleted over large distances. 

Josephson currents are carried solely by the mobile electrons and quantum mechanically subject to the mentioned microscopic tunnelling. {It is of course clear that the strengths of the tunnelling and the tunnelling current depend on the width $L_n$ of the wall because the electron wave function $\psi(x)$ decays with distance. It therefore will be strongest for microscopically thin junction walls which will put us into the high wave number range of the turbulence close to turbulent dissipation.} 

With Josephson frequency $\omega_J=2\pi\nu_J$ the amplitude of the oscillating current density is given by
\begin{equation}\label{eq-jos}
j_n(t)=j_J\sin(\Delta\phi_0-\omega_Jt), \qquad j_J=\frac{e\hbar N_0}{m_eL_n}|\psi|^2
\end{equation}
where $\Delta\phi_0=(\phi_2-\phi_1)_0$ is the original undisturbed phase difference between the interacting magnetic voids, and $|\psi|^2=\delta N/N_0$ the semi-classical fractional density fluctuation $\delta N$ in the interacting nearest neighbour voids normalized to the average density $N_0$. Usually one expects that in the average $\delta N/N_0\sim O(0.1)$ in well developed turbulence. 

The oscillating current $j_n(t)$ is a localized high frequency source (antenna) which necessarily will serve as a radiator of electromagnetic waves. Its oscillation frequency is fairly high depending on the electric potential difference $\Delta V$ which is applied to the junction. Hence it maps even very small potential differences into radiation. One does not expect very strong radiation emitted from a single junction because the current amplitude is of the order of 
\begin{equation}
|j_n|\sim 10^{-23} |\delta N/N_0|N_0L_n^{-1}~~\mathrm{Am}^{-2}\nonumber
\end{equation}
which is inversely proportional to the width $L_n$ of the junction. Radiation from a single current is weak. For an idea chose the magnetosheath. Assuming a density of, say, $N_0\approx 10^3$ m$^{-3}$, a minimum width of the wall $L_n\approx 10^4$ m, and  density contrast of $|\delta N/N_0|\sim 10^{-1}$,  then the single-junction current density amounts to not more than $|j_n|\sim 10^{-25}$ Am$^{-2}$, completely independent on the potential drop $\Delta V$ which enters through phase invariance only. This will contribute just to very weak radiation only. However in  huge volumes, like those of extended objects in the universe, the radiation of all the myriads of junctions present may possibly add up to a measurable intensity. In the following we investigate this possibility.\footnote{{Quantum effects are ignored when $L_n\gg r_0\sim 10^{-10}$ m exceeds the atomic radius $r_0$.  $L_n$ enters the Josephson current, while the gauge invariant quantum phase and $\nu_J$ remain unaffected. The Josephson effect thus exists always in any junction, if $\Delta V\neq0$ across the junction, even though $L_n$ suppresses the current amplitude. Phase gauge invariance in such exceptional cases of spontaneous symmetry breaking, or Berry's phase \citep{kato1950,berry1984} causes observable effects. However high temperatures would in addition obscure these quantum effects. The Josephson frequency is sufficiently far above any reasonable plasma oscillation frequency, and thus spectrally immune against all collisionless temperatures in question. Very high temperatures would destruct any junction inhibiting the Josephson effect. As long as walls and junctions exist (for example in mirror modes in the magnetosheath at temperature $T\sim$ few 10  eV), the Josephson effect will be unavoidable though for small numbers of junctions undetectable, unless a SQUID is used to monitor the Josephson frequency.}}

\subsection{Thermal Josephson  frequency}
Prior to attacking the radiation problem we infer about the expected frequency range of Josephson oscillations. The current  depends on the density contrast $|j_n|\propto |\psi|^2$, while the frequency depends solely on the applied potential difference $V$. Hence the frequency decouples from the radiation problem which can be considered separately later. Since there are no obvious sources expected in the medium to contribute to the potential other than two, the mean thermal fluctuation level of plasma oscillations and plasma turbulence, the question is, which potential differences can they cause? Clearly large and medium potential drops imply high photon energies and accordingly low radiation power. 

Electrostatic thermal fluctuations cause an average rms potential the order of
\begin{eqnarray}
\langle\Delta V\rangle_\mathit{th}&\approx&\ell_D \Big(\frac{2T_e}{\epsilon_0\lambda_D}\Big)^\frac{1}{2}\approx\\
1.24\,\ell_D&\times&10^{-4}\mathrm{V}\:\sqrt[4]{T_{e[\mathrm{keV}]}N_{0[10^3\mathrm{m}^{-3}]}}\nonumber
\end{eqnarray}
Here the dominant length scale of thermal fluctuations is the Debye length $\lambda_D=(2T_e/m_e\omega_e^2)^\frac{1}{2}$. The quantity $\ell_D=L_n/\lambda_D$, a rather large number   is the mean cross-scale number of Debye-lengths over which the potential is measured across the junction boundary (realistic values should be of the order of at least $\ell_D\sim 10^5$ for a junction). Here the electron temperature $T_e\sim$ keV, and a density $N_0\sim 10^{3}$ m$^{-3}$ are used, with density usually taken for the intracluster gas, for instance, inferred basically from X-ray observations under the assumption of fully virialized motion.This thus holds for the rather high-energy component of matter which barely participates in the turbulence as one would expect that the latter involves the denser low-energy part which is less involved into the supposed virial equilibration.  

Hence, even though we assumed a rather high temperature for the intracluster gas based on x-ray observations and the assumption of complete virialization, decreasing the frequency below $\sim1$ GHz into the domain of radio frequency observation requires unrealistically low temperatures (at least for a virialized X-ray cluster). Any radiation will be in the optical to x-ray range.\footnote{Temperatures inferred from X ray observations are very high. It they are the true collisional plasma temperatures, no junction survives them, making our model obsolete.}

It is therefore hardly believable that thermal fluctuations play any remarkable role in generating Josephson radiation, at least not in the radio wavelength, and it will be of very low power. Nevertheless the possibility cannot be excluded that a finite thermal fluctuation level, which cannot be avoided in particular at the high assumed cluster temperatures inferred from X ray observations, contributes to weak high frequency/high energy radiation.

\section{Turbulent Josephson spectrum}
{Returning to the problem of generation of potential drops, we consider the} presence of well developed turbulence. Extended astrophysical objects (stellar winds, SNR, clusters, galaxies) are generally turbulent. The electric field is caused by the turbulent flow, in particular in the presence of an ideal conductivity which on the micro-scales under consideration is always given in extended objects. This applies to almost all candidates (SNRs, the intra-cluster medium etc.) on scales down from  alfv\'enic into the ion-inertial range covering the  interval $\lambda_e< \lambda\lesssim v_A\tau_A$, where $\lambda_e=c/\omega_e\gg\lambda_D$ is the electron skin depth, $\lambda_i=c/\omega_i$ the ion skin depth, $v_A=B/\sqrt{\mu_0N_0m_i}$ the Alfv\'en speed in the mean magnetic field $B\equiv\langle B\rangle$, and $\tau_A$ the Alfv\'en time, a loosely defined quantity only which corresponds to the injection range of mechanical energy into plasma turbulence. The turbulent electric field $\delta\mathbf{E}$ in the scale range $>\lambda_i$ is obtained  from 
\begin{equation}\label{eq-efield}
\delta\mathbf{E}=\langle\mathbf{B}\rangle\times\delta\mathbf{v}-\langle\mathbf{v}\rangle\times\delta\mathbf{B}-\delta\mathbf{v}\times\delta\mathbf{B}+\langle\delta\mathbf{v}\times\delta\mathbf{B}\rangle
\end{equation}
We are less interested in the macroscopic motion $\langle\mathbf{v}\rangle$ of matter and for simplicity drop the second term in this expression. In fact  $\langle\mathbf{v}\rangle$ is basically the macroscopic turbulent rotation speed in the mean magnetic field $\langle\mathbf{B}\rangle$, and the turbulent velocities $\delta\mathbf{v}$ belong to smaller scale  vortices which produce small-scale electric fields responsible for the potential drops across the junctions. The average macroscopic rotation speed $|\langle\mathbf{v}\rangle|=\langle v\rangle_\phi$ can become large in a larger radial range of a  cluster, in particular in its outer skirts where it, however, is about constant and will not contribute remarkably to the potential in a weakly turbulent magnetic field. {Equation (\ref{eq-efield}) refers to the single fluid MHD model of turbulence \citep[cf., e.g.,][]{biskamp2003} and should be refined to a two fluid or kinetic model when leaving the alfv\'enic range, which for our perspective purposes is not required. This point will be discussed in passing below.\footnote{{To preclude any misunderstanding, the present investigation does not contribute to turbulence theory. It merely makes use of turbulence as a model providing the potential drop in Josephson junctions. For economical reasons it restricts itself to the well-established stationary spectral Kolmogorov model in wave number space \citep{kolmogorov1941} which, within wide margins, covers the basic physics in collisionless plasma turbulence. We do not refer to any of its moderate refinements which in the past eighty years since its invention have been constructed in theory as well as from observations. The interested reader is referred to the cited selected literature \citep[cf., e.g.][for review of older work]{biskamp2003}.}}}

\subsection{Potentials}
The {nonlinear} third term and it averaged fourth term in (\ref{eq-efield}) are the correlations between the turbulent velocity and magnetic fields.This fourth terms is driving the turbulent dynamo. To first order we neglect it and the third term, assuming the product of the velocity and field fluctuation amplitudes $|\delta v||\delta B|$ is of second order only, a point which must in a more extended treatment be re-evaluated {but is not of principal importance here}.  Its inclusion would provide an interesting coupling between dynamo and Josephson effects, {which if no resonance occurs is of  second order and thus negligible}. 

The last condition  suggests that the dominant turbulent electric field $\delta\mathbf{E}$ is perpendicular to the mean magnetic and turbulent velocity fields. To first order it is only the mechanical turbulent flow $\delta\mathbf{v}$ that enters the Josephson effect. Since $\langle\mathbf{B}\rangle$ in the wall of the junction is tangential to the junction, the direction of the electric field projects onto the normal $\mathbf{n}$ of the junction with angle $\alpha_k$ depending on the turbulent wave number $\mathbf{k}$. 

Turbulence is conventionally described by reference to the turbulent power spectrum $S(k)$ taken under stationary conditions, with $k$ the modulus of the turbulent wave number in one dimension. Following the usual approach to turbulence, $k$ is the projection of the  wavenumber $\mathbf{k}$ of the turbulent fluctuations onto the direction of the electric field $\delta\mathbf{E}$, the only direction of relevance to our problem which makes it locally unidimensional \citep[cf., e.g.,][]{biskamp2003}. This dimension in our case is along the normal $\mathbf{n}$ of the junction across the wall between the two superconducting voids to which the turbulent electric field $\mathbf{E}$ is projected. The mean magnetic field $\langle\mathbf{B}\rangle$ is confined to the junction wall and thus tangential to the voids forming a magnetic filament. As usual we take this direction as coordinate $z$, and the direction of the normal $\mathbf{n}$ across the junction as $x$. In cylindrical coordinates integration over the angular dependence in the $(y,z)$-plane yields a factor $2\pi$. With $\delta\mathbf{E}(\mathbf{x})=-\nabla\delta V(\mathbf{x})$ we then have\footnote{Strictly speaking the potential of the induction electric field is $V=-\oint \delta\mathbf{E\cdot}d\mathbf{s}$ which yields $\delta V/\delta\mathbf{s}=-\delta\mathbf{E}$, and $\mathbf{s}$ projects onto the normal $\mathbf{n}$ to the narrow wall of the junction. That part then is the potential difference across the junction which we write as the above product with $\mathbf{n\cdot}\nabla=\mathbf{n\cdot}\partial/\partial\mathbf{s}$.}  the potential difference $\Delta V(k)=L_n\mathbf{n\cdot}\delta\mathbf{E}(k)$ across the junction for its spectrum that 
\begin{eqnarray}
S_{E_n}(k)&=&2\pi\langle B\rangle^2S_v(k)\cos^2\alpha_k\\
& \Downarrow & \nonumber\\ 
S_V(k)&=&2\pi L_n^2\langle B\rangle^2S_v(k)\cos^2\alpha_k 
\end{eqnarray}
with $k_\mathit{in}\sim k_A<k<k_m\sim\lambda_i^{-1}$, and angle $\alpha_k$, a function of turbulent wave number $k$, between electric field and direction of the normal. (Indices refer to the different turbulent fields.) {Restriction to scales longer than $\lambda_i$ warrants that for the present time we remain in the alfv\'enic range only.} In this scale range it is known that the turbulent velocity spectrum $S_v(k)$ is about Kolmogorov,\footnote{A Kolmogorov spectrum is sufficiently general for our purposes here. If restriction is made to mhd or anisotropic turbulence, reference to Iroshnikov-Kraichnan \citep{iroshnikov1963} or Goldreich-Shridar \citep{goldreich1997} turbulence, respectively, would be appropriate.} and we have for the spectrum of the potential difference
\begin{equation}
S_V(k) \approx C\langle B\rangle^2\epsilon^\frac{2}{3}L_n^2k^{-\frac{5}{3}}\cos^2\alpha_k
\end{equation}
where $C\approx 10.37$ is Kolmogorov's constant $C_K\approx 1.65$ (modified by the factor $2\pi$) as follows from numerical simulations \citep{kaneda1991,kaneda1993,fung1992,gotoh2001} and \citep[cf., e.g., ][]{biskamp2003}, and $\epsilon$ is the stationary energy injection rate per unit mass and time. {Due to our  selection of terms in Eq.(\ref{eq-efield}) only the velocity spectrum contributes. Inference about the contribution of the magnetic fluctuations through the neglected second convective term in (\ref{eq-efield}) requires consideration of the spectrum of turbulent currents, which is second order and thus outside our purposes.}

Our interest up till now is primarily in the determination of the scale range of relevance for generation of (possibly observable) radiation. Application to the Josephson frequency requires finding the power spectrum of the frequency. Since the frequency is a real quantity and there is no obvious damping\footnote{Josephson oscillations are not damped by themselves. Their decay time is the physical decay time of the junctions, i.e. the decay time of the turbulent magnetic vortices. In stationary turbulence this plays no role because a decaying junction is replaced by some other newly formed one which might have a slightly different Josephson frequency. One thus expects that the bandwidth of the Josephson emissions will be determined mainly by the turbulent fluctuations of scales which is implicitly taken care of in the assumption of the stationary Kolmogorov spectrum.} of the Josephson oscillations on the radiation time scale, this is done by taking the Fourier transform of $\nu_J^2$ which gives directly
\begin{eqnarray}
S_\nu(k)&=&\frac{e^2}{4\pi^2\hbar^2}S_V(k)\\
&=&\frac{e^2L_n^2}{4\pi^2\hbar^2}C\langle B\rangle^2\epsilon^\frac{2}{3}k^{-\frac{5}{3}}\cos^2\alpha_k\nonumber
\end{eqnarray}
If taking the integral with respect to $k$, we find that in the average it turns out that apparently the smallest wave numbers $k_A$ (longest scales) contribute most to the frequency, i.e. they are responsible for the highest frequencies, with an angular average $\langle\cos^2\alpha_k\rangle=\frac{1}{2}$ yielding a spectral average of the turbulent Josephson frequency:
\begin{eqnarray}
\langle\nu_J(k)\rangle&\approx& \frac{eL_n\langle B\rangle\epsilon^\frac{1}{3}}{8\pi\hbar}\big({3C}\big)^\frac{1}{2}k_A^{-\frac{1}{3}}\times\nonumber\\
&\times&\Big[1-\Big(\frac{k_A}{k_m}\Big)^\frac{2}{3}\Big]^\frac{1}{2}|\cos\alpha_k|\\
&\approx&3\times10^{4}L_n\epsilon^\frac{1}{3}\langle B\rangle k_A^{-\frac{1}{3}}|\cos\alpha_k|~~\mathrm{GHz} \nonumber
\end{eqnarray}
Here  $ k_A^{-1}=v_A\tau_A \approx L_{J\|}$ is roughly the length of the junction, and we integrated over the angle $\alpha_k$. (Note the dimension of $[\epsilon]=\mathrm{m}^2/\mathrm{s}^3$.) All these quantities are to be taken locally in the generation region of radiation, the nearest neighbour interaction of the junctions. So the conclusion that the longest scales contribute most should be cautioned, because only a fraction of the long scales maps to the junction walls. The effective scale responsible for contributing to the Josephson frequency is the product $kL_n$.

\subsection{Wave number dependence}
More interesting than the average frequency is the power spectrum of the Josephson frequency as function of junction-projected scale $k$. It holds in the alfv\'enic range, i.e. the long wavelength range $k_A<k<k_m$ with frequency per root wavenumber as function of the turbulent scale  
\begin{eqnarray}
\frac{\nu_J(k)}{\sqrt{k}}&=&\frac{\sqrt{C}}{2\pi}\frac{e}{\hbar}L_n\langle B\rangle \epsilon^\frac{1}{3} k^{-\frac{5}{6}}|\cos\alpha_k|\\\nonumber
&\approx&10^{6}L_n\langle B\rangle \epsilon^\frac{1}{3} k^{-\frac{5}{6}}|\cos\alpha_k| ~~\mathrm{GHz\ m}^{\frac{1}{2}}
\end{eqnarray}
Note that the physical SI units in the second part of this equation and the following two estimates have been absorbed. This implies the unit $[B]=$ m$^{-2}$ here. The magnetic field is measured in flux elements $\hbar/e$ here. 

It is worth mentioning that $L_nL_y\langle B\rangle$ is the flux contained in the wall carried by the magnetic filament separating the junction voids, if $L_y$ is the dimension of the wall perpendicular to the main magnetic field. Multiplying with $\Phi_0^{-1}= e/\pi\hbar$, the inverse elementary flux, gives the number $N_\Phi\approx L_nL_y\langle B\rangle$ of flux elements the wall contains which participate in the Josephson oscillation. Its large number is the reason for the high Josephson frequency. 

The Josephson frequency depends on the turbulent scale as  $\nu_J(k)\propto k^{-\frac{1}{3}}$ GHz, weakly decaying towards the larger turbulent wave numbers. Smaller turbulent wave numbers $k$ contribute to higher Josephson frequencies. Writing this in terms of the electron skin depth $\lambda_e=c/\omega_e$ gives
\begin{equation}
{\nu_J(k)}\approx10^{6}\lambda_e^{\frac{1}{3}}L_n\langle B\rangle \epsilon^\frac{1}{3} (k\lambda_e)^{-\frac{1}{3}}|\cos\alpha_k| ~~\mathrm{GHz}
\end{equation}

This expression depends on the projection angle $\alpha_k$ the turbulent electric fields make with the normal to the junction. Large angles $\alpha_k$ reduce the potential thus lowering the Josephson frequency. The largest potentials and thus frequencies are obtained for angle $\alpha_k=0$. Turbulent flow velocities nearly parallel to the normal $\delta\mathbf{v}\approx \mathbf{n}$ contribute to the lowest frequencies because $\mathbf{E}$ is perpendicular to $\mathbf{n}$ for such flow directions. The direction of turbulent flow with respect to the junction (and magnetic field) thus modulates the Josephson frequency over a wide frequency range from $\nu_J\approx 0$ to its maximum, which is reached at $k=k_A$ and $\alpha_{k_A}=0$. 

It is convenient to introduce the ratio $\kappa=k/k_A$ and writing for $k_A\approx \omega_{ci}/v_A=\omega_i/c\equiv\lambda_i^{-1}$ where $\omega_{ci}=e\langle B\rangle/m_i$ is the ion cyclotron frequency. Then $k_A\lambda_e\approx \sqrt{m_e/m_i}$. For protons the above expression can, with $\ell_n=L_n/\lambda_e$,  be re-written more conveniently as
\begin{eqnarray}\label{nu-fin}
\nu_J(\kappa)&=&\nu_{JA}|\sin\beta_k|\kappa^{-\frac{1}{3}}\\
 \nu_{JA}&\approx&10^6\ell_n\lambda_e^\frac{4}{3}\langle B\rangle c^\frac{2}{3}\Big({\epsilon/c^2}\Big)^\frac{1}{3}~~\mathrm{GHz}
\end{eqnarray}
  Here $\alpha_k=\pi/2-\beta_k$ has been replaced by its complementary angle $\beta_k$. Following the above discussion, small $\beta_k\ll \pi/2$ are responsible for low Josephson frequencies. However, the frequency will remain high, which is seen when using the above given numbers. Josephson oscillations, once mapping into escaping radiation, should provide diffuse high-frequency/photon energy radiation backgrounds of extended turbulent objects independent on the presence of particles that have been accelerated to high energies. 
  
\begin{figure}[t!]
\centerline{\includegraphics[width=0.5\textwidth,clip=]{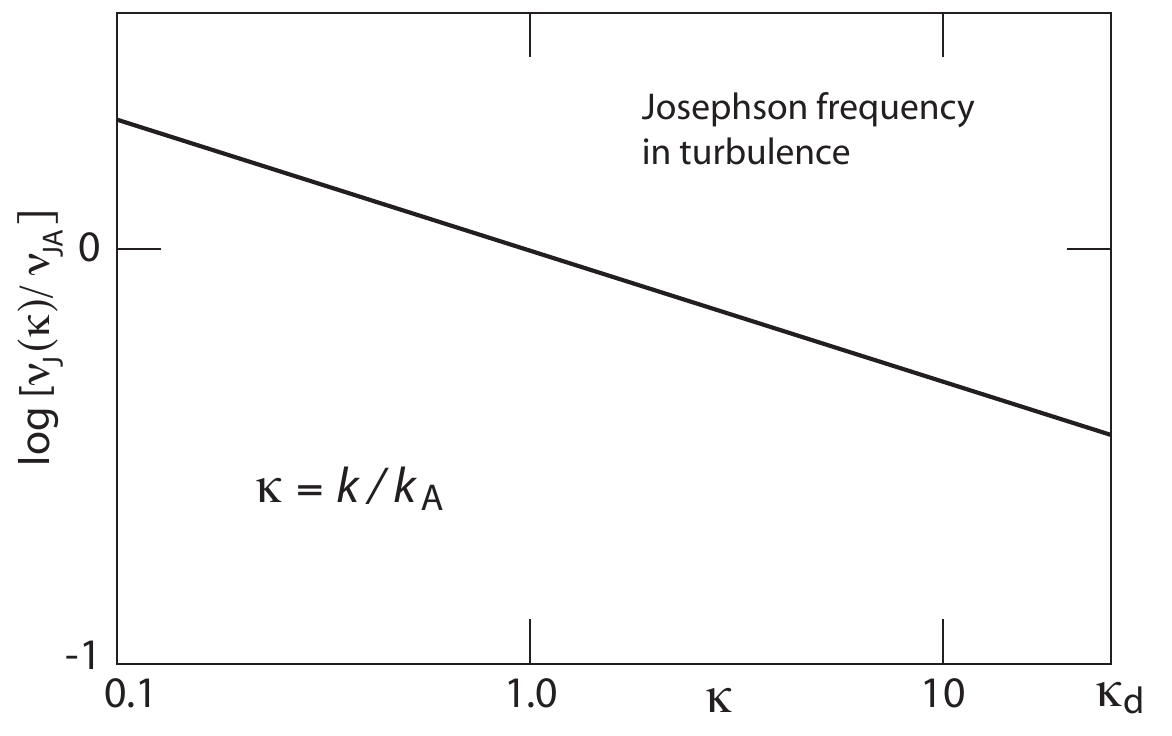}}
\caption{Dependence of the normalized Josephson frequency in Kolmogorov turbulence on the ratio $\kappa\equiv k/k_A$ of turbulent wave numbers in the range $0.1 k_A<k<k_d$ under the assumption that $k_d\lambda_e\approx 1$ is the ultimate dissipation scale of turbulence which is still far above the molecular scale, caused for instance by spontaneous collisionless reconnection in inertial scale turbulent current filaments, the most probable dissipation process in turbulence in collisionless plasma \citep[cf., e.g.][]{treumann2015}.} \label{josep}
\end{figure}

The dependence of the Josephson frequency on the normalized wave number $\kappa$ in Kolmogorov turbulence is shown in Figure \ref{josep} up to the wavenumber of dissipation $\kappa_d$ which corresponds to scales where {particle inertial effects become} important. The decay of frequency with increasing $\kappa$ amounts to roughly 1.5 orders of magnitude only, however showing that large turbulent wave numbers near dissipation contribute most to the lowest frequencies.  

{The coefficient of the alfv\'enic frequency in the above expression corresponds to an oscillation energy of $\approx 4$ eV, which directly maps into the frequency of radiation. The junction width will be $\ell_n>10$, while $\lambda_e\sim 10^4$ m. Let us assume $B\sim 1$ nT somewhat larger than the magnetic field in the outskirts of clusters of galaxies. Then $\nu_J\sim 2\times 10^{6}(\epsilon/c^2)^{1/3}$ Hz, depending on the square of the mechanical turbulent velocity injected per second. Assume $v/c\sim 10^{-5}$ corresponding to $v\sim 10^3$ km/s. Then largest frequencies are $\nu_J\sim 4\times10^4$ Hz for $\beta_k=\pi/2$, in the low frequency radio range. Depending on the angle, the Josephson frequency covers the range from zero frequency to this maximum. Frequencies  below the local radiation cut off belong to quasi-stationary Josephson current-generated non-dynamo magnetic fields.}

\subsection{Ion inertial range effect}
We briefly discuss the contribution of the ion-inertial wave number range $\lambda_i>k^{-1}>\lambda_e$. Ions become non-magnetic there, {and Eq. (\ref{eq-efield}) is to be re-interpreted in the sense of electron-MHD \citep[cf., e.g.,][]{gordeev1994,lyutikov2013}. This leads to deviations from alfv\'enic magnetic turbulence with Alfv\'en waves turning into kinetic Alfv\'en waves of transverse scale $k_\perp\lambda_i\sim 1$, parallel wave electric fields \citep{lysak1996}, and inclusion of shorter scale $k\lambda_e<1$ whistler turbulence \citep{lyutikov2013}. Parallel electric fields do not contribute to the Josephson frequency $\nu_J$. A wealth of observations indicate that the turbulent magnetic spectra deviate from Kolmogorov in this range \citep[cf., e.g.,][and the literature cited therein]{bale2005,alexandrova2009,alexandrova2020,alexandrova2021,matteini2017,breuillard2018,stawarz2021}.\footnote{{Such deviations, if markedly affecting the velocity spectrum, modify the Josephson frequency spectrum on ion scales. This requires a more precise theory and model of $S_v(k)$ than Kolmogorov. (Some MMS electron observations seem to suggest that the electron spectra $S_e(k)$ parallels the turbulent spectrum of the electric field $S_E(k)$ which is somewhat flatter than Kolmogorov \citep{stawarz2016,gershman2018} in this range. Indeed, the currents are carried by electrons and the electric fields are due to current instabilities, mostly kinetic Alfv\'en waves on these scales. Thus the relation between $S_e(k)\sim S_E(k)$ is reasonable. It does however not justify the conclusion that the mechanical turbulence $S_v(k)\sim S_e(k)$ follows the electrons.) Thus for being cautious and conservative, we interpret the spectral cut-off $k_d$ accordingly that, if extension of Kolmogorov's $S_v(k)$ into the ion range is not warranted by observations, then $k_d\lambda_i\sim1$, and those wave numbers are understood as already belonging into the dissipative range. They then do neither contribute to Josephson frequency nor radiation.}}   As long as no bulk flow is included, any magnetic turbulence does to first order not contribute to Josephson oscillations and radiation. In view of application to the Josephson effect this justifies extension of the active part of the turbulent spectrum down into the ion-inertial range at wave number $k_d$ where dissipation sets on.\footnote{{Briefly leaving our main route, we comment on two sorts of observations ($a$) in the ion inertial and ($b$) in the dissipative wave number ranges. $(a)$ Occasionally large electric field amplitudes have locally been measured \citep[cf., e.g.,][]{bale2005}. Whatsoever the reason is for their generation (which for our purposes is of little interest but should be attributed to nonlinear, i.e. higher order, effects like electron holes or small-scale shocklets etc.), and if the junction concept can be maintained (i.e. presence of magnetic vortices and voids separated by walls) such average potentials would shift $\nu_J$ up into the X ray domain though with presumably unobservable current (and radiation) intensity. Singular Josephson effects are unobservable. They could, however, be detected putting a SQUID on the spacecraft to catch the Josephson signal, measuring the electric field with high precision.  ($b$) Observations in the dissipation region (by whatsoever process dissipation is caused) exhibit either exponential or algebraic spectral cut-offs in magnetic turbulence interpreted by different dissipation models of magnetic energy. This range corresponds to our high wavenumber cut-off $k_d$ beyond which no junctions will evolve as the dissipation range is not anymore turbulent, cutting the Josephson frequency sharply at $k_d$; no other contribution to the turbulent Josephson effect is expected from larger wave numbers.}}}   

There is, however another contribution to the electric potential in this range which comes from the ion response to the turbulent induction electric field in the ion inertial range 
$k\lambda_i>1$ and is responsible for deviations of the turbulent density spectrum from its original Kolmogorov shape \citep{treumann2019}.   This is of higher order and can be neglected.

{Turbulent flow and currents in the ion inertial range are restricted to electrons. The relation between the turbulent electric and velocity fields includes Poisson's equation.} The divergence of the electric field $\delta E$ is obtained from (\ref{eq-efield}) being proportional to the turbulent density fluctuation $\delta N$. It causes a correction on the spectral density of the electric potential field
\begin{equation}
S_{V,N}(k)=\frac{e^2}{\epsilon_0^2}k^{-2}S_N(k)
\end{equation}
 caused by $S_N(k)$, the power spectral density of the turbulent density $\delta N$. The latter  is proportional to the power spectrum $S_v(k)$ of the turbulent velocity
\begin{equation}
S_N(k)\approx N_0^2\frac{v_A^2}{c^2}\frac{k^2}{\omega_i^2}S_v(k)
\end{equation}
Note again that nothing is changed on the turbulent spectrum $S_v(k)$ of the mechanical velocity which is imposed on the electromagnetic fluctuations. We therefore have 
\begin{equation}
S_{V,N}(k)\propto S_v(k)
\end{equation}
The factor $k^2$ is compensated by the required factor in the Fourier transformed Poisson equation. {(It might be noted that this expression could, in principle, be refined if taking into account an active response of the plasma through the inclusion of the inverse plasma response function $D^{-1}(\omega,\mathbf{k})$, which would generate a more complicated dependence of the density power spectrum for comparison with observation.)}  

The effect is of second order in $v_A/c$ and thus weak. Its contribution can to first order be suppressed. It just affects the density spectrum to let it response to the electric field \citep[for its reconstruction in real observations see][]{treumann2019} which it experiences as a charge field. More important is that in this range the turbulent velocity is determined by electron mobility. Electron-MHD takes to some extent care of it. The magnetic turbulence spectrum decays steeper than Kolmogorov \citep{lyutikov2013}. Thus to first order, neither magnetic effects (for principal reasons) nor density fluctuations are of interest in the turbulent Josephson effect.

\section{Radiation}
The single junction oscillates in the whole spectral range with Josephson frequency $\nu_J(k)$ as function of wave number $k$ or scale $l\sim k^{-1}$. The oscillating Josephson current $j_n(t)$ acts as a current source for the generation of escaping radiation. According to the above the Josephson frequency and thus the radiation frequency is far above any plasma cut-off $\omega_J\gg\omega_e$ and thus can freely escape into space such that we do not need to consider any radiation transport or reabsorption as in all applications the matter will locally be optically thin while of course over large spatial scales might be subject to scattering on the material background medium.

The total energy $dW$ radiated into solid angle $d\Omega$ is defined as 
\begin{equation}
\frac{dW}{d\Omega}={\textstyle\frac{1}{2}}\epsilon_0c^3\int|\mathbf{A}(t)|^2dt={\textstyle\frac{1}{2}}\epsilon_0c^3\int|\mathbf{A}(\omega)|^2d\omega
\end{equation}
where $\mathbf{A}$ is the radiated vector potential, $\omega$ is the radiation frequency. Since no single junction will be resolved in the volume, the expected radiation of all the many junctions to which we refer below is diffuse for any remote observer, and one may integrate over the  solid angle even for a single junction in order to obtain the radiation intensity per frequency interval  
\begin{equation}
\frac{dI}{d\omega}\approx {4\pi\epsilon_0 c^3}\sum_\pm|\mathbf{A}(\pm\omega)|^2~~~\frac{\mathrm{J}}{\mathrm{Hz}}
\end{equation}
\begin{figure}[t!]

\centerline{\includegraphics[width=0.5\textwidth,clip=]{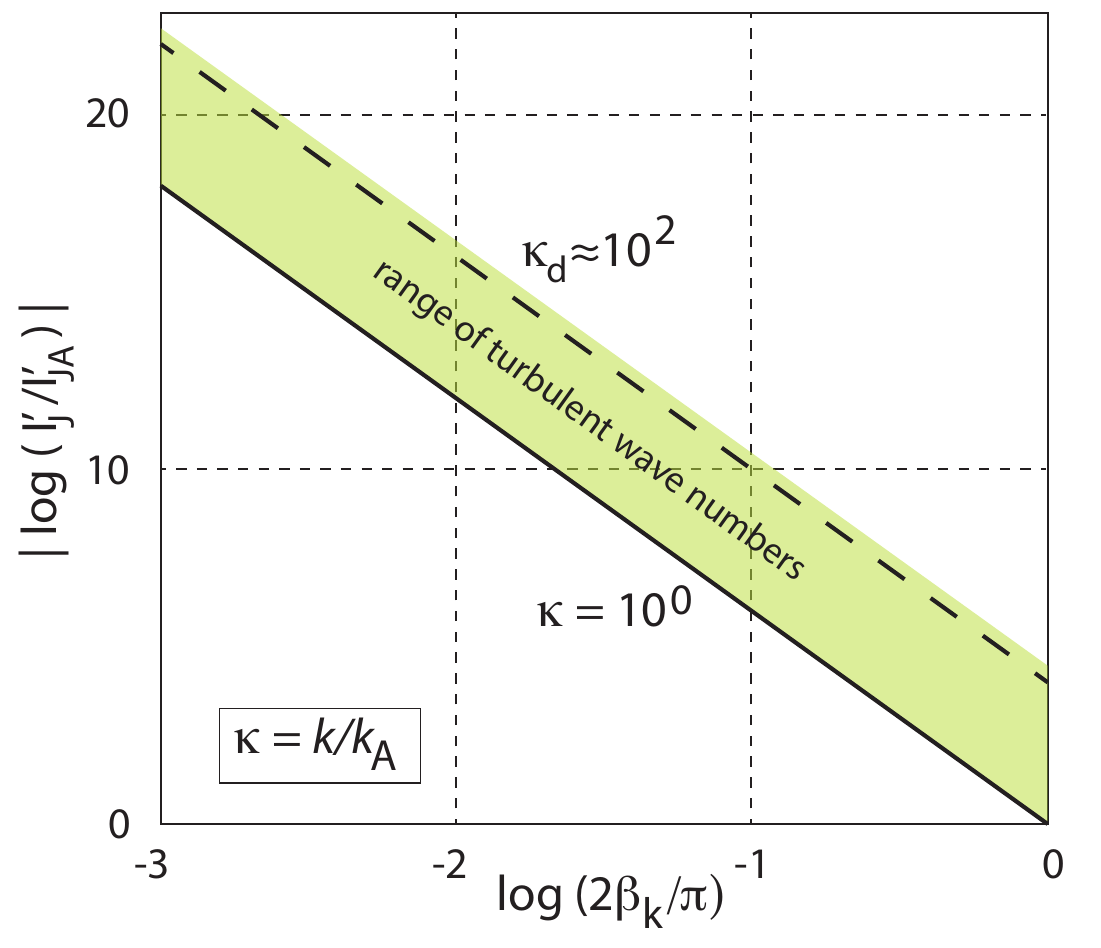}}
\caption{Dependence of the normalized  intensity per frequency $(dI/d\omega_J)/(dI/d\omega_{JA})$ on the decrease of the angle $10^{-3}\leq\frac{2}{\pi}\beta_k\leq1$ for two ratios $\kappa\equiv k/k_A$ of turbulent wave numbers in the range $0.1 k_A<k<k_d$ under the assumption that $k_d\lambda_e\approx 1$ is the turbulent dissipation scale and turbulent energy injection is at or below the alfv\'enic scale $k_A^{-1}$. The ultimate turbulent dissipation scale $k_d^{-1}$ is assumed to be two orders apart at $\kappa\approx 10^2$. The interval between the two lines indicates the spectral width of the turbulence in its contribution to Josephson radiation (shown in green colour). The radiation intensity increases  drastically with decreasing angle $\beta_k$, i.e. with the obliqueness of the turbulent electric field with respect to the normal to the junction.  } \label{josep-1}
\end{figure}

\subsection{Vector potential}
It remains to determine the radiation vector potential as function of frequency, 
solving the inhomogeneous wave equation with Josephson current as source. The wave equation for the remaining component $A_x$ is  
\begin{eqnarray}
\Big[\nabla^2-\frac{\partial^2}{c^2\partial t^2}\Big]\frac{{A}_x(\mathbf{x},t)}{\mu_0}&=&-j_x(\mathbf{x},t)\\
&&j_J\sin(\omega_Jt)\delta(\mathbf{x-x}_0)\nonumber
\end{eqnarray}
with Josephson current in direction $\mathbf{n}=\mathbf{\hat x}$. Here $\mathbf{x}_0$ is the location of the junction in real space, and the spatial dependence of the Josephson frequency is suppressed. For any observer the junction is a point  source taken approximately rectangular neglecting any curvature effects. Also the stationary phase difference $\Delta\phi_{\,0}$ is of no importance here as it drops out when calculating the radiation intensity. The magnetic field is in $z$-direction, the radiation wave vector $\mathbf{K}$  varies in $y,z$. For an order of magnitude estimate  avoiding the complete Greens function solution (which  implies retaining just the dipolar term \citep{jackson1975,rybicki1979}, in which case the source current becomes a plane current with plane extended over short distance along the magnetic field in $z$ and perpendicular in $y$), we seek for a particular solution of the wave equation. Fourier transformation with respect to time yields 
\begin{eqnarray}
&&\big[\nabla^2+\frac{\omega^2}{c^2}\big]A_x(\mathbf{x},\omega)=\\
&&{i\pi\mu_0j_J}\big[\delta(\omega+\omega_J)-\delta(\omega-\omega_J)\big]\delta(\mathbf{x-x}_0)\nonumber
\end{eqnarray} 
Note that the radiation wavelength at the high frequencies is of course much shorter than any turbulence scale which justifies the neglect of the dependence on the turbulent scale $\mathbf{x}'=(y,z)$. Physically this implies that the energy loss by radiation is negligible against the turbulent energy. Fourier transforming yields 
\begin{eqnarray}
A_x(\mathbf{K},\omega)&=&- \frac{2\pi^2i\mu_0j_J\,\mathrm{e}^{-iK\mathbf{n\cdot x}_0}}{K^2-\omega_J^2/c^2}\\
&\times&\big[\delta(\omega+\omega_J)-\delta(\omega-\omega_J)\big]\delta(\mathbf{x-x}_0)\nonumber
\end{eqnarray}
with $\mathbf{x}_0$ the location of the junction in real space somewhere in the huge turbulent volume. The singularity in the denominator has to be treated  accounting for  causality of outgoing radiation requires  $K=\pm\omega_J/c$.  Resolving the nominator yields 
\begin{eqnarray}
A_x(\mathbf{K},\omega)&=&-\frac{\pi^3\mu_0cj_J\,\mathrm{e}^{-iK\mathbf{n\cdot x}_0}}{\omega_J}\big[\delta(\omega+\omega_J)-\delta(\omega-\omega_J)\big]\nonumber\\[-2ex]
\\[-1.5ex]
&\times&\big[\delta(K+\omega_J/c)+\delta(K-\omega_J/c)\big]\delta(\mathbf{x-x}_0)\nonumber
\end{eqnarray}
For a single turbulent junction the amplitude becomes 
\begin{equation}
|A_x(\omega)|_{\omega=\omega_J}=4\pi^3\mu_0c^2j_J\omega_J^{-3}
\end{equation}

\subsection{Intensity}
The energy emitted per frequency $\omega=\omega_J$ in radiation is
\begin{eqnarray}
\frac{dI}{d\omega}\Big|_{\omega_J}&\approx& \frac{64\pi^7 c^3}{\epsilon_0\omega^6_J}j_J^2\\
&=&\frac{64\pi^5}{\mu_0c}\Big(\frac{\lambda_e}{L_n}\Big)^2\Big(\frac{\omega_e}{\omega_J}\Big)^6\Big(\frac{\langle\delta N\rangle}{N_0}\Big)^2\Phi_0^2\nonumber\\
&\approx& 2.1\times10^{-28}\ell^{-2}_n\Big(\frac{\omega_e}{\omega_J}\Big)^6\Big(\frac{\langle\delta N\rangle}{N_0}\Big)^2~\frac{\mathrm{J}}{\mathrm{Hz}}\nonumber
\end{eqnarray}
We introduced the electron skin depth $\lambda_e=c/\omega_e$, plasma frequency $\omega_e^2=e^2N_0/\epsilon_0m_e$, flux quantum $\Phi_0=\pi\hbar/e$, the ratio $\ell_n=L_n/\lambda_e$ and $\langle\delta N\rangle$, the rms turbulent density fluctuation.  
This  depends sensitively on the Josephson frequency. For the large Josephson frequencies the intensity of a single junction becomes very low. Inserting the expression (\ref{nu-fin}) with $\omega_e=2\pi\nu_e$ one obtains
\begin{eqnarray}
\frac{dI}{d\omega}\Big|_{\omega_J}&\approx& 1.3\times10^{-27}\Big(\frac{\kappa}{\ell_n}\Big)^2\Big(\frac{\nu_e}{\nu_{JA}}\Big)^6\Big(\frac{\langle\delta N\rangle}{N_0}\Big)^2\nonumber\\
&\approx&\frac{6.6\times10^{-57}\kappa^2}{\ell_{n}^{8}\nu_e^2\epsilon^2\langle B\rangle^6|\sin\beta_k|^6}
\end{eqnarray}
Assuming $\ell_n\sim O(10^2)$, $\nu_e\sim 10^4$ Hz, and taking for the magnetic field $\langle B\rangle\sim 1$ nT, produces for one single junction an emitted intensity per Hz of
\begin{equation}
\frac{dI}{d\omega}\Big|_{\omega=\omega_J}\approx \frac{6.6\times10^{-20}\kappa^2}{\ell^8_{n[10]}\nu_{e[10^4]}^2\langle B\rangle^6_{[nT]}\epsilon^2|\sin\beta_k|^6}~\frac{\mathrm{J}}{\mathrm{Hz}}
\end{equation}
which corresponds roughly to some $10^{-8}$ eV/Hz. This expression is very sensitive to the strength of the ambient magnetic field.  This value increases slightly with the participating wave number range $\kappa$ and decays with the rate of energy injection into  turbulence, which is basically unknown. The increase with $\kappa$ is at most some factor $10^2$ at the turbulent dissipation range. There is, however, a rather sensitive dependence on the angle $\beta_k$ as we already discussed when dealing with the frequency. For small $\beta_k$ implying lowest frequencies, the radiated intensity increases as the sixth power of $\beta_k$. This makes, for instance, for an angle $\beta_k\sim 0.1\pi/2$ that the intensity increases by a factor of $\sim10^5$. If the angle is $\beta_k\sim 0.01\pi/2$ the intensity is increased by a factor $\sim10^{11}$. Figure \ref{josep-1} sketches the dependence of the emitted intensity as function of $\beta_k$ for two different spectral ranges $\kappa$.

 The largest spectral contribution comes from the large turbulent wavenumber range near dissipation which also provides the lowest radiation frequencies. This suggests that the  part of the spectrum of turbulence near dissipation wave numbers $k_d$ contributes most to possibly observable radiation. {Collisionless turbulence at those scales, still far away from any molecular interaction, is believed to dissipate its magnetic energy in the ion $k\lambda_i\gtrsim1$ and electron $k\lambda_e\gtrsim1$ inertial ranges by nonlinear plasma processes (see footnote 8)}, one concludes that any observation of low-frequency Josephson radiation in the large-scale structure of the universe is probably related to the direct signature of the wave number $k_d$ above that collisionless turbulence enters its ultimate dissipation region.

\subsection{Volume filling factor}
 Locating the oblique turbulent vortices into the large wave number turbulent range near dissipation then yields that the emitted intensity in radiation comes close to the eV-range in energy per Hz and per $\epsilon^2$. Though this value remains rather slow in particular for high energy injection rates and stronger ambient magnetic fields, it shows that the short wave number range of well developed turbulence can indeed provide radiation at the Josephson frequency.  However, one single junction will in general not generate any susceptible radiation which could be measured from remote, in particular not if the radiation source is at cosmological distance with radiation intensity decaying inversely proportional to some power of the cosmological red shift. 

Observations never deal with one single junction which for any objects in the universe is of microscopic size and thus undetectable. Any of the large turbulent volumes available in the universe will however contain a large number of different junctions distributed over the entire volume. A precise calculation requires knowledge of their spatial distribution function and the solution of the dipolar radiation pattern for each of the junction which we so far avoided. Since the distribution is not known, a proxi to the sum over all  contributions of junctions is provided by the estimate of the volume filling factor of the microscopic junctions and multiplying the radiation. The volume filling factor is defined as 
\begin{equation}
\xi=\sum_s p_{Js}\mathcal{V}_0/\langle\Delta\mathcal{V}\rangle_{Js}
\end{equation}
 where $p_{Js}<1$ is the normalized probability of encountering a turbulent Josephson junction in the volume $\mathcal{V}_0$  and $\langle\Delta\mathcal{V}\rangle_{Js}$ is the average junction volume. This factor will be large, even for small  probabilities. For  a junction volume $\langle\mathcal{V}\rangle_{Js}\sim 10^{16}$ m$^3$ and a spherical turbulence volume $\mathcal{V}_0\sim3\times10^{64}$ m$^3$ with probability not more than a mere $p_{Js}\sim 10^{-10}$,  the average filling factor is $\langle\xi\rangle\sim10^{37}$, which increases the total average radiation intensity, replacing the numerical factor in the above expression by a factor $\sim 10^{10}$.

\section{Conclusions}
In this brief communication we dealt with an unusual effect in classical high-temperature collisionless media as those encountered in  extended astronomical objects like SNRs and clusters of galaxies which have evolved into a state of quasi-stationary turbulence in an extended range $k_A\lesssim k\lesssim k_d$ of turbulent wave numbers $k$. On the  junction scales the matter is a dilute and collisionless ideally conducting plasma. In its turbulent state it consists of a texture of vortices of different scales which at the assumed elevated temperatures are current vortices. They  evolve into a texture of magnetic voids surrounded by magnetic filaments. Examples can be found in the large-scale structure of the universe which is known to exhibit a particular filamentary texture. We assume that this can be understood as a set of grossly independent Josephson junctions consisting each of two adjacent voids and the separating magnetic wall whose width is sufficiently narrow for permitting electron tunnelling between the voids. Nearest neighbour interactions dominate. Josephson conditions suggest that these junctions, which are penetrated by smaller scale vortices causing small cross-junction potential drops, emit radiation at Josephson frequency. The  power emitted by one single junction is tiny, but filling the volume up with many such similar junctions adds up to possibly observable radiation intensity. 

The main contribution to radiation intensity is provided by the smallest-scale vortices near dissipation in developed turbulence.  The dissipation scale does not contribute, however,  because it lacks any junctions. In general Josephson radiation is at low frequency. It will be observed only if the frequency exceeds the ambient plasma frequency cut off. This may be the case in dilute large astrophysical objects like the outskirts of galaxy clusters where huge volume filling factors raise its intensity. Part of its energy is then deposited into weak diffuse radiation in the radio range via the myriads of tiny junctions that form in the course of turbulence. The spectral range is, from the above estimates, of the order of one decade. 

Josephson spectra extend down to zero frequency, which implies that  Josephson currents are sources of stationary magnetic fields. They generate a magnetic texture independent of any dynamo action. 

In all cases a mechanism is needed that produces a cross junction electric potential drop $\Delta V$ however weak. This is most probably provided by turbulence. For obtaining any measurable power in field or radiation, it requires a large volume providing large filling factors $\xi$.

\end{document}